\documentclass[conference]{IEEEtran}
\IEEEoverridecommandlockouts

\usepackage{cite}
\usepackage{amsmath,amssymb,amsfonts}
\usepackage{algorithmic}
\usepackage{graphicx}
\usepackage{textcomp}
\usepackage{xcolor}
\usepackage[euler]{textgreek}
\usepackage{color}
\usepackage{url}
\usepackage{todonotes}
\usepackage{multirow}
\usepackage{tabularx}
\usepackage{subfig}
\usepackage{enumitem}

\begin{document}

\title{Modelling Fog Offloading Performance}

\author{\IEEEauthorblockN{Ayesha Abdul Majeed, Peter Kilpatrick, Ivor Spence, and Blesson Varghese}
\IEEEauthorblockA{\textit{School of Electronics, Electrical Engineering and Computer Science, Queen's University Belfast, UK}\\
E-mail: \{aabdulmajeed01, p.kilpatrick, i.spence, b.varghese\}@qub.ac.uk}
}

\maketitle

\thispagestyle{plain}
\pagestyle{plain}

\begin{abstract}
Fog computing has emerged as a computing paradigm aimed at addressing the issues of latency, bandwidth and privacy when mobile devices are communicating with remote cloud services. The concept is to offload compute services closer to the data. However many challenges exist in the realisation of this approach. During offloading, (part of) the application underpinned by the services may be unavailable, which the user will experience as down time. This paper describes work aimed at building models to allow prediction of such down time based on metrics (operational data) of the underlying and surrounding infrastructure. Such prediction would be invaluable in the context of automated Fog offloading and adaptive decision making in Fog orchestration. Models that cater for four container-based stateless and stateful offload techniques, namely Save and Load, Export and Import, Push and Pull and Live Migration, are built using four (linear and non-linear) regression techniques. Experimental results comprising over 42 million data points from multiple lab-based Fog infrastructure are presented. The results highlight that reasonably accurate predictions (measured by the coefficient of determination for regression models, mean absolute percentage error, and mean absolute error) may be obtained when considering 25 metrics relevant to the infrastructure. 
\end{abstract}


\section{Introduction}
\label{sec:introduction}
\begin{figure}[t]
    \centering
    \includegraphics[width=0.4\textwidth]{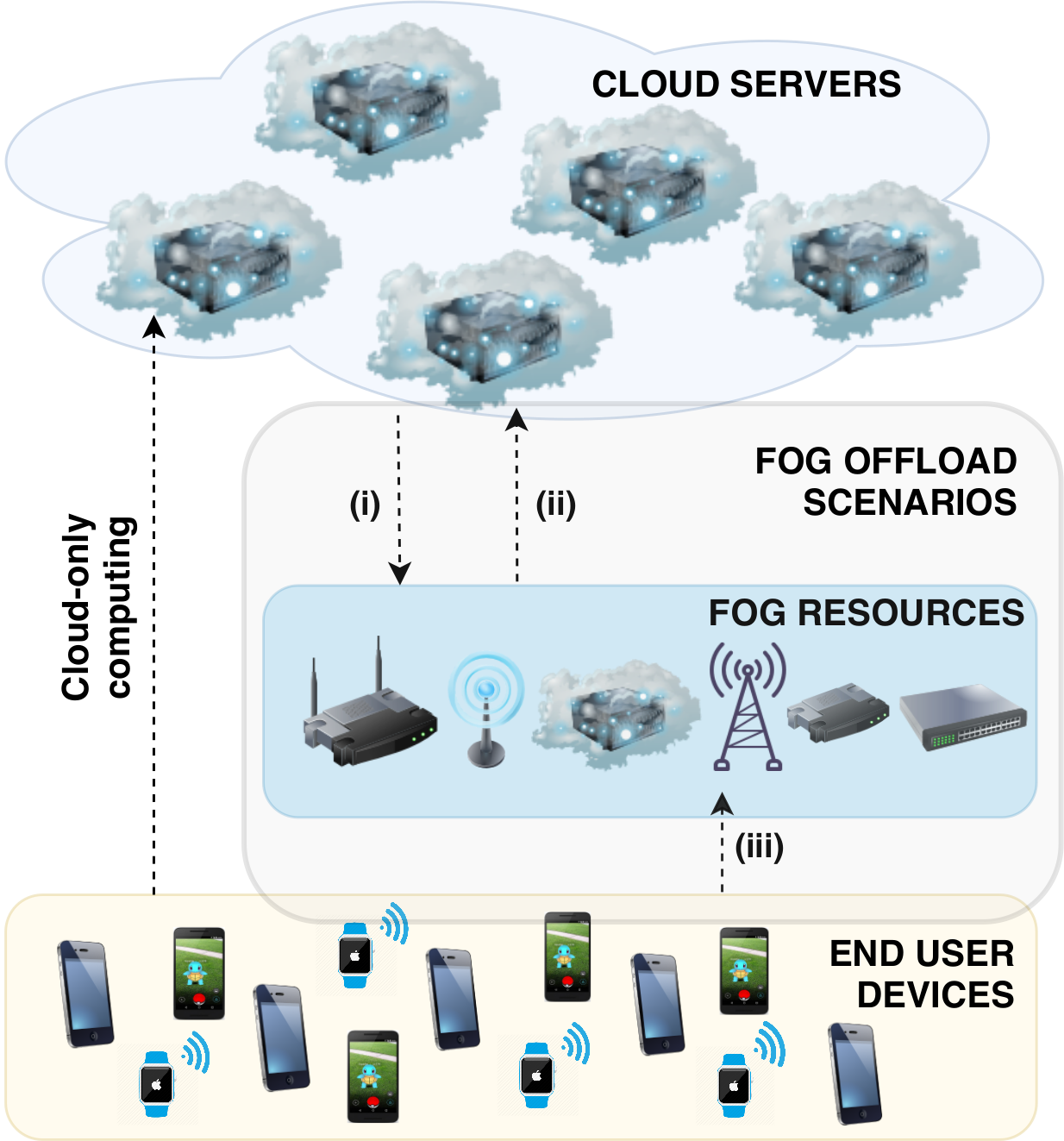}
    \caption{Potential Fog offload scenarios, namely (i) Cloud-to-Fog, (ii) Fog-to-Cloud, and (iii) Device-to-Fog}
    \label{fig:offloading}
\end{figure}

The limited resources of mobile devices, both power and computational, led to the Cloud-only concept of offloading services to a Cloud data centre. A
desire to reduce the latency introduced by such offloading, together with an increase in the availability of resources (for example
micro data centres or compute enabled routers) at the edge of the network, has resulted in the more flexible Fog computing approach.
As illustrated in Figure~\ref{fig:offloading}, services can be offloaded: (i) from the Cloud to the Fog, to minimise the latency
of communicating with the end-user device~\cite{VARGHESE2018849}, (ii) from Fog to the Cloud, to exploit additional
computational and storage resources~\cite{management2018}, and (iii) from end-user devices to the Fog, to satisfy compute requirements unavailable on the devices~\cite{osmoticcomputing-01}. We consider Cloud-to-Fog and Fog-to-Cloud offloading.

The benefits derived from offloading~\cite{satyanarayanan2017emergence} can be negated if the very process of offloading results
in a temporary loss of service (down time). We aim to predict the times that would be taken to offload a given service by a variety of known methods, which are
estimates of the corresponding induced down times. Such predictions would be helpful in the following situations:

(i) \textit{Automated Fog software development environments}: predictive models would enable a Fog software developer to assess
the feasibility of offloading without establishing corresponding testbeds and conducting empirical
investigations.

(ii) \textit{Adaptive decision-making in Fog orchestration platforms}: the transient nature of Fog environments means
that the best selection of services, destinations and techniques for offloading may well be constantly changing. Predictive models
would permit a continual reassessment of offloading decisions.

(iii) \textit{Simulation platforms}: in the absence of corresponding computational and network infrastructure, researchers and practitioners can
use simulation models to predict the performance of a particular configuration. Predictive models of offload times would make the models
more complete in that down times as well as runtime performance could be modelled.

The paper aims to address the following \textit{three research questions} - \textbf{Q1}: How can different offloading approaches be modelled? \textbf{Q2}: What runtime (system and network related) and offline (do not change during offload process) parameters influence offloading? \textbf{Q3}: Given the same empirical data of offloading from different experimental platforms how is the accuracy of estimation affected when using different estimation methods and machine learning algorithms?

In this investigation, we consider three `Stateless' techniques (Save and Load, Export and Import, and Push and Pull) and one `Stateful' technique (Checkpoint/Restore in Userspace), which are container-based techniques adopted in this paper for offloading from the Cloud to the Fog and vice-versa. As a first step, the anatomy of each of the offloading approaches is considered by studying the individual components that make up the overall time required to offload (addresses Q1). The research establishes a method for estimating the time taken to offload a service which will also be applicable for different offloading techniques. 25 relevant runtime and offline metrics for the entire Cloud and Fog system have been identified to model the offloading time (addresses Q2). Four machine learning based predictive models (one linear and three non-linear) are applied to over 42 million data points obtained from two experimental platforms (to address Q3). 
The results indicate that the chosen runtime and offline metrics are relevant to predicting the offload time which can be estimated with reasonable accuracy.

This paper makes the following \textit{three research contributions}: (i) Develops a method for modelling the offloading time for multiple container-based stateless and stateful techniques, (ii) Identifies 25 runtime and offline parameters that influence the model for offloading, and (iii) Presents an empirical investigation of the accuracy of multiple linear and non-linear machine learning algorithms for predicting the offloading time. 

The remainder of this paper is organised as follows. 
Section~\ref{sec:techniques} discusses stateless and stateful techniques for offloading.
Section~\ref{sec:estimationmodels} proposes two methods for estimating the performance of offloading and highlights the models for estimating performance.
Section~\ref{sec:experiments} presents the experimental studies pursued on two different platforms. 
Section~\ref{sec:relatedwork} highlights research relevant to the discussion of this paper. 
Section~\ref{sec:conclusions} concludes this paper by considering future work.

\section{Fog Offloading}
\label{sec:techniques}
This section presents three offloading scenarios highlighted in Figure~\ref{fig:offloading} and four container-based offloading techniques.

\subsection{Offloading Scenarios} 
Fog offloading can be considered in the following three scenarios, which are illustrated in Figure~\ref{fig:offloading}: 

(i) \textit{Cloud-to-Fog offloading}~\cite{wang2017enorm},\cite{mcchesney2019defog} - this refers to transferring services of an application from a Cloud resource on to a Fog resource to meet latency and ingress bandwidth demands. Since the offloaded service is closer to the device that generates data, it reduces the communication latency and (pre)-processes data closer to the source, thereby reducing the volume of data that needs to be transferred to the cloud. 

(ii) \textit{Fog-to-Cloud offloading}~\cite{wang2017enorm, popularitybased} - this refers to transferring the services that are resident on a Fog resource to the Cloud. This may be done in response to a change in the life-cycle of the application service - the service on the Fog resource needs to be terminated and resumed on the Cloud from where it was originally offloaded. This may be because the service requires additional resources (CPU cores, storage, memory) that are not available on the Fog resource, but can only be satisfied on the Cloud. 

(iii) \textit{Device-to-Fog offloading}~\cite{osmoticcomputing-01, yousefpour2018reducing} - this refers to transferring services of an application from one or a collection of end user devices or sensors to a Fog resource. This is done in order to preserve battery life of devices or to meet the computational demands of workloads that cannot be executed on user devices due to limited form factor and weak processing capabilities. The offloaded service executes on the Fog resource and the resulting output is provided back to the devices or sensors. 

The focus of this paper is the Cloud-to-Fog and Fog-to-Cloud offloading scenarios.

\subsection{Container-Based Offloading Approaches}
There are two dominant approaches that facilitate offloading. The first is Virtual Machine (VM) migration. VM hand-off is used to move a service for supporting user mobility in the context of cloudlets~\cite{ha2015adaptive}. A synthesis technique is adopted in which the VM is divided into two stacked overlays to optimise the downtime during VM hand-off. 

A second approach to facilitate offloading is container migration. Containers are an alternative virtualisation approach that are lightweight and portable, thereby making offloading quicker than using VMs~\cite{soltesz2007container}. Therefore, containers are investigated within Fog computing, in which Fog resources have limited resources when compared to the Cloud~\cite{wang2017enorm,management2018}. Popular container technologies, include LXC\footnote{\url{https://linuxcontainers.org/}} and Docker\footnote{\url{https://www.docker.com/}} and they support migration, which is required for offloading. 

Fog offloading, in this paper, is explored in the context of Docker containers. Docker packages an application as an image that consist of a file system with the required libraries, executables and configuration files. In practice, the image may comprise a series of layers that are stacked on top of a base image, for example, the Ubuntu operating system. When a container is executed Docker mounts all the layers of the image as `read-only' using the Union File System (UnionFS) and the top layer as a writable layer as shown in Figure~\ref{fig:dockercontaineroffloading}.

Docker supports two migration approaches that can be used for offloading: (i) \textit{Stateless}~-~the state of the application is not transferred, instead a separate instance of the container is run elsewhere without its previous state (for example, Save and Load, Export and Import, and Push and Pull), and (ii) \textit{Stateful}~-~the state of the running application is transferred with the container image. Stateful migration can be achieved by using the CRIU (Checkpoint/Restore in Userspace)\footnote{\url{https://criu.org/Main_Page}} approach.

\subsubsection{Stateless Techniques}
Three stateless techniques are considered in this paper for offloading. They are as follows: 
\subsubsection*{(i) Save and Load}
\begin{figure}[ht]
    \centering
    \includegraphics[width=0.5\textwidth]{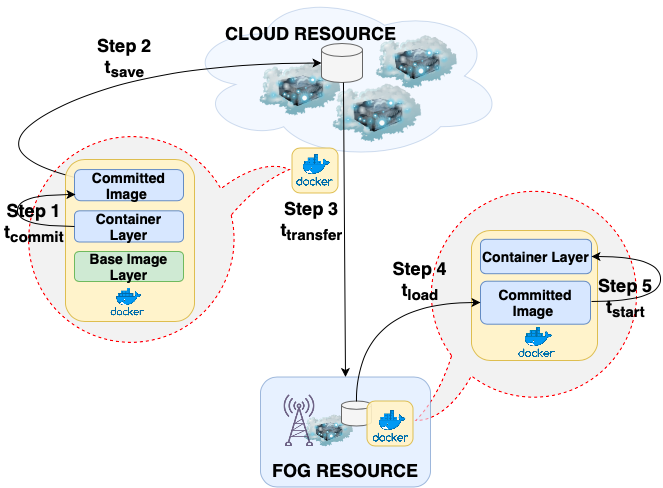}
    \caption{Five steps in offloading a container from the Cloud to the Fog using the Save and Load technique}
    \label{fig:dockercontaineroffloading}
\end{figure}

Figure~\ref{fig:dockercontaineroffloading} shows the Cloud-to-Fog offloading scenario using the Save and Load technique. 
The goal of this technique is to transfer the image of a running container and the underlying base layers. This offloading technique is a five step process as illustrated below: 

\textit{Step 1 - Commit}: A container is instantiated when it boots up from a series of base image layers 
when there is a request for offloading from the Cloud server. The commit operation stops the running container and saves the current state of the container, the accompanying stacked layers, and the modifications that were made within the container as a new image, referred to as the committed image. This newly created image is stored in the local image registry of the Cloud server. The time taken for this step (store the configuration and run-time state of the container, create an image, and store the image in the local registry) is denoted as $t_{commit}$.

\textit{Step 2 - Save}: The save operation, converts the committed image to a compressed \texttt{.tar} file and saves it on the hard disk of the Cloud server. The compressed file contains information on the contents of the container, including its parent layers and the size of each layer. The time taken to convert the committed image to the compressed file is denoted as $t_{save}$. 

\textit{Step 3 - Transfer}: The compressed image file is transferred to the Fog resource using a network transfer protocol, such as FTP, SCP or rsync; $t_{transfer}$  captures the time taken to transfer the compressed file to the Fog server.

\textit{Step 4 - Load}: This operation initially decompresses the \texttt{.tar} image file and loads it from the hard disk on the Fog resource as an image into the local image registry. This time is captured as $t_{load}$.

\textit{Step 5 - Start}: A container from the image in the registry is booted up on the Fog resource; $t_{start}$ captures this time.

Based on the above five steps, the total time taken to offload a container-based service from the Cloud to the Fog or vice-versa is represented as:
\begin{equation}
t_{offload} = t_{commit} + t_{save} + t_{transfer} + t_{load} + t_{start}
\label{eqn:toffload}
\end{equation}

In the Cloud-to-Fog offload scenario,
$t_{commit}$ and $t_{save}$ are the operations on the Cloud server, and $t_{load}$ and $t_{start}$ are on the Fog resource. The transfer time will be for transferring the compressed file from the Cloud to the Fog. In the Fog-to-Cloud offload scenario, the commit and save times will be for the operation on the Fog resource, and the load and start for the operations on the Cloud server. 

In this work it is assumed that Fog applications are designed and developed as micro-services (a collection of services can be orchestrated as a workflow and each service can be deployed as a container). The advantage of this offloading technique is that the base image layers are also transferred from the source (for example, the Cloud server in the Cloud-to-Fog offload scenario) to the destination Fog resource allowing for creation of further copies of the container. However, there will be a trade-off with the size of the image transferred. 

A potential advantage is when multiple containers that rely on the same underlying libraries need to be offloaded. In this case, the base image layers will not be duplicated for multiple container services when it is offloaded. In addition, the history of the image is preserved (configuration settings, ports and entry points), which is an advantage, but incurs a large performance overhead during container deployment.

\begin{figure}[ht]
    \centering
    \includegraphics[width=0.5\textwidth]{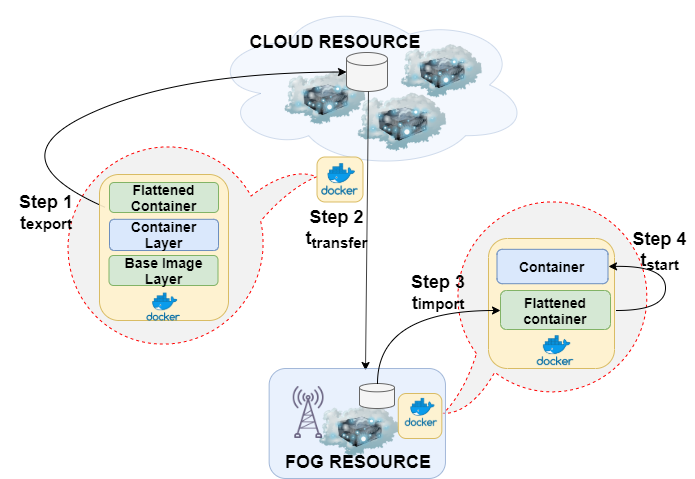}
    \caption{Four steps in offloading a container from the Cloud to the Fog using the Export and Import technique}
    \label{fig:offloadingexport}
 \end{figure}

\subsubsection*{(ii) Export and Import}
Figure~\ref{fig:offloadingexport} shows the Cloud-to-Fog offloading scenario using the Export and Import technique. The goal is to export the running state as a flattened (multiple image layers into one single layer) container. This offloading technique is a four step method as illustrated below: 

\textit{Step 1 - Export}: The export operation stops the running container and saves the current state of the container and the modifications that were made within the container, referred to as  the flattened container. The flattened container is stored on hard disk of the Cloud server as a \texttt{.tar} file.
The time taken for this step (store the run-time state of the container, and store the container as a tar file) is denoted as $t_{export}$. 

\textit{Step 2 - Transfer}: The compressed container file is transferred to the Fog resource using SCP network transfer protocol. $t_{transfer}$ time captures the time taken to transfer the compressed file to the Fog server.

\textit{Step 3 - Import}: This operation initially decompresses the \texttt{.tar} file and imports it from the hard disk on the Fog resource as an image into local image registry. This time is captured as $t_{import}$.

\textit{Step 4 - Start}: A container is booted up on the Fog resource from the flattened container in the registry; this time is $t_{start}$.

The total time taken to offload is represented as:
\begin{equation}
t_{offload} = t_{export} +  t_{transfer} + t_{import} + t_{start}
\label{eqn:toffloadimport}
\end{equation}

Flattening the container removes the history (configuration settings, ports and entry point settings) of the container which reduces the container size and improves the efficiency of the deployment process of a container.
One disadvantage of this approach is that the flattened container that is exported cannot be used as a template to create a new container.

\subsubsection*{(iii) Push and Pull Technique}
\begin{figure}[ht]
    \centering
    \includegraphics[width=0.5\textwidth]{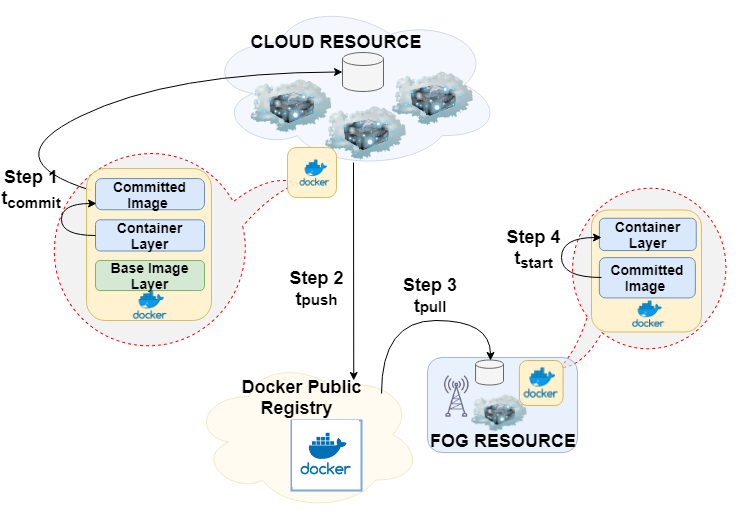}
    \caption{Four steps in offloading a container from the Cloud to the Fog using the Push and Pull technique}
    \label{fig:offloadingpullandpush}
\end{figure}

Figure~\ref{fig:offloadingpullandpush} shows the Cloud-to-Fog offloading scenario using the Push and Pull technique - the image of a running container is pushed from the Cloud server to a central repository. Then the image is pulled and a container is instantiated on the Fog server.

This  technique is a four step method as illustrated below: 

\textit{Step 1 - Commit}: The commit operation saves the current state of the container, and the modifications that were made within the container. The committed image is stored on the local registry of the Cloud server.
The time taken for this step (store the configuration and run-time state of the container, and store the container in the local registry) is denoted as $t_{commit}$. 

\textit{Step 2 - Push}: The committed image is pushed into the Docker Hub using push command, $t_{push}$ captures the time taken to push the committed image into the Docker hub.

\textit{Step 3- Pull}: This operation pulls the image from the Docker hub and loads the image into the local image registry of the Fog server. This time is captured as $t_{Pull}$.

\textit{Step 4 - Start}: A container from the image in the registry is booted up on the Fog resource; $t_{start}$ captures this time.

The total time taken to offload a container-based service from the Cloud to the Fog or vice-versa is represented as:
\begin{equation}
t_{offload} = t_{commit} +  t_{push} + t_{pull} + t_{start}
\label{eqn:toffloadpullnpush}
\end{equation}

When the available network bandwidth is limited, pushing and pulling an image from the Docker Hub slows container deployment. Docker repository is a data-intensive application and the repository becomes a performance bottleneck, as the number of images and user requests increases, thereby affecting the offloading process.

\subsubsection{Stateful Techniques}
A container-based live migration technique is presented in this paper. 

\subsubsection*{Checkpoint/Restore in Userspace (Live Migration)}
Figure 5 shows the Cloud-to-Fog offloading scenario using the live migration technique. Docker uses Checkpoint/Restore in Userspace (CRIU) to freeze a running container by checkpointing it and then restoring it. The process is as follows:

\textit{Step 1 - Checkpoint}: This procedure freezes the container process, collects and saves the complete state of the container and then stops the container ($t_{checkpoint}$ captures this time). The container state file (dump files) is shared (or transferred) to the Fog server through shared Network File System (NFS).

\textit{Step 2 - Restore}: a container is restored from the dump files on the mounted directory, process execution is resumed (this time is $t_{restore}$). Container is initialised after it is restored on the Fog server. The offload time is represented as:
\begin{equation}
t_{offload} = t_{checkpoint} +  t_{restore} 
\label{eqn:toffloadcriu}
\end{equation}

\begin{figure}[ht]
    \centering
    \includegraphics[width=0.5\textwidth]{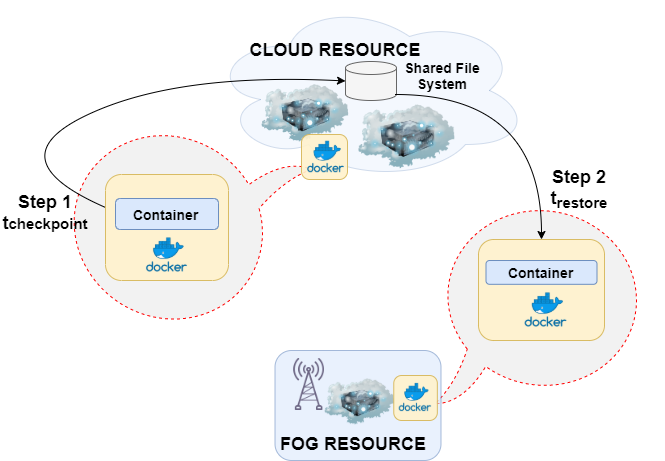}
    \caption{Two steps in offloading a container from the Cloud to the Fog using the Stateful offloading technique}
    \label{fig:criu}
\end{figure}

Stateful techniques are beneficial for applications in which a client is serviced by an application that must retain the internal state of the application. One disadvantage is that services cannot be further replicated on a different Fog node since data relevant to individual users is stored in the application.


\section{Estimation Methods and Models}
\label{sec:estimationmodels}
In this section, the parameters that influence the stateless and stateful offloading techniques, the methods used for estimation, and the machine learning algorithms used for the estimation model that predicts $t_{offload}$ are presented.

\subsection{Methods for Estimation}
Table~\ref{tab:tab1} describes a catalogue of parameters that are considered in this work by the estimation models while offloading a container. Two types of parameters are considered, namely runtime and offline parameters. The \textit{runtime} parameters are collected when the steps of the migration techniques are executed to offload a container. These parameters relate to both the offloading process and the entire Cloud and/or Fog system as highlighted in the table. The network properties between the Cloud and the Fog are considered to capture the state of the network when the offload occurs. The \textit{offline} parameters are statically determined and do not change during the offload process; for example number of cores, network bandwidth and container image size.

\begin{table}[t]
\caption{Parameters that impact overall offloading time; Bps - Bytes per second, bps - bits per second, BW - bandwidth}
\label{tab:tab1}
\vspace{-3pt}
\begin{tabular}{|l|l|c|l|}
\hline
\textbf{Parameter} & \textbf{Description} & \multicolumn{1}{l|}{\textbf{\begin{tabular}[c]{@{}c@{}}System/\\Process\end{tabular}}} & \textbf{\begin{tabular}[c]{@{}c@{}}Cloud/\\Fog\end{tabular}} \\ \hline
\multicolumn{4}{|c|}{\textit{Runtime}} \\ \hline
$P_{1}$, $P_{9}$ & CPU utilisation (\%) & \multirow{3}{*}{System} & \multirow{3}{*}{\begin{tabular}[c]{@{}l@{}}$P_{1}$ - $P_{3}$ (Cloud)\\$P_{9}$ - $P_{11}$ (Fog)\end{tabular}} \\ \cline{1-2}

$P_{2}$, $P_{10}$ & Memory utilisation (\%) &  &  \\ \cline{1-2}
$P_{3}$, $P_{11}$ & Disk utilisation (\%) &  &  \\ \hline
$P_{4}$, $P_{12}$ & CPU utilisation (\%) & \multirow{3}{*}{\begin{tabular}[c]{@{}c@{}}Offloading \\ Process\end{tabular}} & \multirow{3}{*}{\begin{tabular}[c]{@{}l@{}}$P_{4}$ - $P_{8}$ (Cloud)\\$P_{12}$ - $P_{16}$ (Fog)\end{tabular}} \\ \cline{1-2}
$P_{5}$, $P_{13}$ & Memory utilisation (\%) &  &  \\ \cline{1-2}
$P_{6}$, $P_{14}$ & Disk throughput (Bps) &  &  \\ \cline{1-2}
$P_{7}$, $P_{15}$ & Bytes sent (KB/sec) &  &  \\ \cline{1-2}
$P_{8}$, $P_{16}$ & Bytes received (KB/sec)  &  &  \\ \hline
\multicolumn{4}{|c|}{\textit{Offline}} \\ \hline
$P_{17}$ & Image size (MB) & \multicolumn{2}{c|}{\begin{tabular}[c]{@{}c@{}}Offloaded container\end{tabular}} \\ \hline
$P_{18}$, $P_{21}$ & No of Cores & \multicolumn{2}{c|}{\multirow{3}{*}{\begin{tabular}[c]{@{}c@{}}$P_{18}$ - $P_{20}$  (Cloud)\\ $P_{21}$ - $P_{23}$ (Fog)\end{tabular}}} \\ \cline{1-2}
$P_{19}$, $P_{22}$ & Memory Size (GB) & \multicolumn{2}{c|}{} \\ \cline{1-2}
$P_{20}$, $P_{23}$ & Hard disk Size (GB) & \multicolumn{2}{c|}{} \\ \hline
$P_{24}$ & Network BW (bps) & \multicolumn{2}{c|}{\multirow{2}{*}{\begin{tabular}[c]{@{}c@{}}Between \\ Cloud and Fog\end{tabular}}} \\ \cline{1-2}
$P_{25}$ & Network Latency (ms) & \multicolumn{2}{c|}{} \\ \hline
\end{tabular}
\end{table}

\begin{table}[t]
\caption{Parameters that impact the offloading sequence of the Save and Load Technique}
\label{tab:tab2}
\centering
\vspace{-3pt}
\begin{tabular}{| l | l |}
\hline
\multicolumn{1}{|c|}{\textbf{Time}} & \multicolumn{1}{c|}{\textbf{Parameters that impact}} \\ \hline
$t\textsubscript{commit}$ & $X_{commit} = \{$ P\textsubscript{1}, $\cdots$, P\textsubscript{6}, P\textsubscript{17}, $\cdots$, P\textsubscript{20} $\}$ \\
$t\textsubscript{save}$ & \begin{tabular}[c]{@{}l@{}}$X_{save} = \{$ P\textsubscript{1}, $\cdots$, P\textsubscript{6}, P\textsubscript{17},$\cdots$, P\textsubscript{20}\end{tabular} \}\\
$t\textsubscript{transfer}$ & $X_{transfer} = \{$ P\textsubscript{7}, P\textsubscript{8}, P\textsubscript{15}, P\textsubscript{16}, P\textsubscript{17}, P\textsubscript{24}, P\textsubscript{25} $\}$\\
$t\textsubscript{load}$ & $X_{load} = \{$ \begin{tabular}[c]{@{}l@{}}P\textsubscript{9}, $\cdots$ P\textsubscript{14}, P\textsubscript{17}, P\textsubscript{21}, $\cdots$ P\textsubscript{23}$\}$\end{tabular} \\
$t\textsubscript{start}$ & \begin{tabular}[c]{@{}l@{}}$X_{start} = \{$ P\textsubscript{9}, $\cdots$, P\textsubscript{14}, P\textsubscript{17}, P\textsubscript{21}, $\cdots$, P\textsubscript{23}$\}$\end{tabular} \\ \hline
\end{tabular}
\end{table}
\begin{table}[t]
\caption{Parameters that impact the offloading sequence of the Export and Import Technique}
\label{tab:tab3}
\centering
\vspace{-3pt}
\begin{tabular}{| l | l |}
\hline
\multicolumn{1}{|c|}{\textbf{Time}} & \multicolumn{1}{c|}{\textbf{Parameters that impact}} \\ \hline
$t\textsubscript{export}$ & $X_{export} = \{$ P\textsubscript{1}, $\cdots$, P\textsubscript{6}, P\textsubscript{17}, $\cdots$, P\textsubscript{20} $\}$ \\
$t\textsubscript{transfer}$ & $X_{transfer} = \{$ P\textsubscript{7}, P\textsubscript{8}, P\textsubscript{15}, P\textsubscript{16},P\textsubscript{17}, P\textsubscript{24}, P\textsubscript{25} $\}$\\
$t\textsubscript{import}$ & $X_{import} = \{$ \begin{tabular}[c]{@{}l@{}}P\textsubscript{9}, $\cdots$ P\textsubscript{14}, P\textsubscript{17}, P\textsubscript{21}, $\cdots$ P\textsubscript{23} $\}$\end{tabular} \\
$t\textsubscript{start}$ & \begin{tabular}[c]{@{}l@{}}$X_{start} = \{$ P\textsubscript{9}, $\cdots$, P\textsubscript{14},P\textsubscript{17} P\textsubscript{21}, $\cdots$, P\textsubscript{23}$\}$\end{tabular} \\ \hline
\end{tabular}
\end{table}
\begin{table}[t]
\caption{Parameters that impact the offloading sequence of the Push and Pull Technique}
\label{tab:tab4}
\centering
\vspace{-3pt}
\begin{tabular}{| l | l |}
\hline
\multicolumn{1}{|c|}{\textbf{Time}} & \multicolumn{1}{c|}{\textbf{Parameters that impact}} \\ \hline
$t\textsubscript{commit}$ & $X_{commit} = \{$ P\textsubscript{1}, $\cdots$, P\textsubscript{6}, P\textsubscript{17}, $\cdots$, P\textsubscript{20} $\}$ \\
$t\textsubscript{push}$ & \begin{tabular}[c]{@{}l@{}}$X_{pull} = \{$ P\textsubscript{1}, $\cdots$, P\textsubscript{8}, P\textsubscript{17}, $\cdots$, P\textsubscript{20},P\textsubscript{24},P\textsubscript{25} \end{tabular} \}\\
$t\textsubscript{pull}$ & $X_{push} = \{$ \begin{tabular}[c]{@{}l@{}}P\textsubscript{9}, $\cdots$, P\textsubscript{17}, P\textsubscript{21}, $\cdots$, P\textsubscript{25} $\}$\end{tabular} \\
$t\textsubscript{start}$ & \begin{tabular}[c]{@{}l@{}}$X_{start} = \{$ P\textsubscript{9}, $\cdots$, P\textsubscript{14}, P\textsubscript{17} $\}$\end{tabular} \\ \hline
\end{tabular}

\end{table}
\begin{table}[t]
\caption{Parameters that impact the offloading sequence of the Stateful Technique}
\label{tab:tab5}
\centering
\vspace{-3pt}
\begin{tabular}{| l | l |}
\hline
\multicolumn{1}{|c|}{\textbf{Time}} & \multicolumn{1}{c|}{\textbf{Parameters that impact}} \\ \hline
$t\textsubscript{checkpoint}$ & $X_{checkpoint} = \{$ P\textsubscript{1}, $\cdots$, P\textsubscript{8}, P\textsubscript{17}, $\cdots$, P\textsubscript{20} $\}$ \\
$t\textsubscript{restore}$ & \begin{tabular}[c]{@{}l@{}}$X_{restore} = \{$ P\textsubscript{9}, $\cdots$, P\textsubscript{16}, P\textsubscript{21}, $\cdots$, P\textsubscript{23} \end{tabular} \}\\
\hline
\end{tabular}
\end{table}

Two methods are used to estimate the offload time - the first uses a single estimation model and the second uses multiple estimation models for the individual time components shown in  Equation~\ref{eqn:toffload}-\ref{eqn:toffloadcriu}.

The first method is \textit{using a collective model}, referred to as (CM), which is a reference to the use of a single model that estimates the offload time. The collective model uses all the parameters listed in Table~\ref{tab:tab1} as input. Consider a collective model, $M_{collective}$ that estimates the offload time and let  $X_{offload} = \{P_{1},\cdots, P_{25}\}$  be the input to the model, then we represent $t_{offload}$ = $M_{collective}(X_{offload})$. 

The second method is \textit{using individual models}, which refers to the use of separate models for estimating the individual times of Equations~\ref{eqn:toffload}-\ref{eqn:toffloadcriu}. 

Table~\ref{tab:tab2} shows the parameters that affect the individual times of the Save and Load technique. Let $M_{commit}$ be an individual model to estimate $t_{commit}$ using the input $X_{commit}$ shown in Table~\ref{tab:tab2}. Then the estimation of $t_{commit}$ = $M_{commit}(X_{commit})$.
Similarly, $t_{save}$ = $M_{save}(X_{save})$, 
$t_{transfer}$ = $M_{transfer}(X_{transfer})$, 
$t_{load}$ = $M_{load}(X_{load})$, and
$t_{start}$ = $M_{start}(X_{start})$.
The offload time can be estimated as shown in Equation~\ref{eqn:individualSave}.
\begin{multline}
\label{eqn:individualSave}
    t_{offload} = M_{commit}(X_{commit})+M_{save}(X_{save})+\\
    M_{transfer}(X_{transfer})+M_{load}(X_{load})+M_{start}(X_{start})
\end{multline} 

Table~\ref{tab:tab3} shows the parameters that affect the individual times of the Export and Import technique. Let $M_{export}$ be an individual model to estimate $t_{export}$ using the input $X_{export}$ shown in Table~\ref{tab:tab3}. Then the estimation of $t_{export}$ = $M_{export}(X_{export})$.
Similarly, $t_{transfer}$ = $M_{transfer}(X_{transfer})$, 
$t_{import}$ = $M_{import}(X_{import})$, and
$t_{start}$ = $M_{start}(X_{start})$.
The offload time can be estimated as shown in Equation~\ref{eqn:individualExport}.
\begin{multline}
\label{eqn:individualExport}
    t_{offload} = M_{export}(X_{export})+M_{transfer}(X_{transfer})+\\
    M_{import}(X_{import})+M_{start}(X_{start})
\end{multline} 

Table~\ref{tab:tab4} shows the parameters that affect the individual times of the Push and Pull technique. Let $M_{commit}$ be an individual model to estimate $t_{commit}$ using the input $X_{commit}$ shown in Table~\ref{tab:tab4}. Then the estimation of $t_{commit}$ = $M_{commit}(X_{commit})$.
Similarly, $t_{pull}$ = $M_{pull}(X_{pull})$, 
$t_{push}$ = $M_{push}(X_{push})$, and
$t_{start}$ = $M_{start}(X_{start})$.
The offload time can be estimated as shown in Equation~\ref{eqn:individualPull}.
\begin{multline}
\label{eqn:individualPull}
    t_{offload} = M_{commit}(X_{commit})+M_{pull}(X_{pull})+\\
    M_{push}(X_{push})+M_{start}(X_{start})
\end{multline} 

Table~\ref{tab:tab5} shows the parameters that affect the individual times of the stateful technique. Let $M_{checkpoint}$ be an individual model to estimate $t_{checkpoint}$ using the input $X_{checkpoint}$ shown in Table~\ref{tab:tab5}. Then the estimation of $t_{checkpoint}$ = $M_{checkpoint}(X_{checkpoint})$.
Similarly, $t_{restore}$ = $M_{restore}(X_{restore})$.
The offload time can be estimated as shown in Equation~\ref{eqn:individualcriu}.
\begin{multline}
\label{eqn:individualcriu}
    t_{offload} = M_{checkpoint}(X_{checkpoint}) + M_{restore}(X_{restore})
\end{multline}

\subsection{Models for Estimation}
\label{subsec:models}
Four machine learning algorithms (one linear and three non-linear) were explored for predicting the offload time. The approach used for estimation is based on historical data that is collected from the experimental platform (presented in Section~\ref{sec:experiments}) to predict $t_{offload}$. The algorithms used are:

(i) Multivariate Linear Regression (MLR): The model developed using this algorithm captures the relationship between multiple input variables $X = \{ P_{1}, P_{2},\cdots, P_{n}\}$ and the dependent output variable $t_{offload}$ by a straight line equation~\cite{pham2017predicting}.

(ii) Polynomial Multivariate Regression (PMR): The regression model developed captures the relationship between the input variables $X = \{ P_{1}, P_{2},\cdots, P_{n}\}$ and the dependent output variable $t_{offload}$ as an $n^{th}$ degree polynomial in $X$.

(iii) Random Forest Regression (RFR): An ensemble model generates $k$ different training subsets from the original data set, and then $k$ different decision trees are built based on the generated training subsets. Each sample of the testing data set is predicted by all decision trees, and the final result is obtained by averaging a score specific to each decision tree~\cite{pham2017predicting}.

(iv) Ridge regression (RR): A non-linear approach that adds an penalty ($L2$), which equals the sum of the squared value of the coefficients $Error_{L2}=Error+\sum_{i=0}^{N}\lambda.{W_{i}}^{2}$~\cite{le1992ridge}.


\section{Experimental Studies}
\label{sec:experiments}
This section presents the experimental setup and the results obtained from running experiments for the Stateful and Stateless container based offloading approaches.

\subsection{Experimental setup}
The proposed methods, namely the Collective Model (CM) and the Individual Models (IM) for estimating the performance of offloading using four estimation models, namely MLR, PMR, RFR and RR are evaluated on two different lab-based platforms (each platform is a combination of a Cloud and Fog VM that executes the container based offloading techniques). Docker 18.09-ce is installed on all VMs.
Both experimental platforms use 64-bit x86 architectures for the Cloud and Fog environment. A number of recent Fog-enabled nodes, such as the Dell Edge Gateway 5000, use 64-bit x86 processors~\cite{puliafito2019fog}. 

The first platform is the combination of a Cloud VM running Ubuntu 18.10 with 6 virtual CPUs, 30GB hard disk and 6GB RAM and Fog VM running Ubuntu 18.10 with 2 virtual CPUs, 20GB hard disk and 2GB RAM.

The network bandwidth between the Cloud and Fog VMs are emulated using Linux Traffic Control (\texttt{tc})\footnote{https://linux.die.net/man/8/tc}. The bandwidth is varied as 25Mbps, 50Mbps, 100Mbps, 1000Mbps with a latency of 10ms and 30ms (values are based on the literature~\cite{ma2017efficient}). 

The second platform is another combination of a Cloud VM and Fog VM running on OpenStack. Both VMs run Ubuntu 18.10; the Cloud VM has 4 virtual CPUs, 80GB hard disk space and 8GB virtual RAM whereas the Fog VM has 2 virtual CPUs, a 40GB hard disk, and 4GB virtual RAM. The default network connection is approximately 3.2Mbps.

\subsubsection{Stateless Techniques}
The following is considered for the set up of the stateless techniques.
For the Cloud VM in the \textit{Save and Load} technique, the container state is saved using the Docker commit process and the save process saves the image as a \texttt{.tar} file. The tar file is transferred to the Fog server using the scp protocol. When the tar file is received on the Fog VM, \texttt{inotify} loads the image into the local registry. A container is started from the loaded image. 
During the offloading process, the values of the run-time parameters are collected at one-second intervals. System CPU utilisation is captured by $P_{1}$ and $P_{9}$, which are obtained by monitoring \texttt{/proc/stat}. CPU utilisation of the offloading process captured by $P_{4}$ and $P_{12}$ are obtained by monitoring \texttt{/proc/PID/}. RAM utilisation of the system, denoted by $P_{2}$ and $P_{10}$ are obtained by the Linux utility tool \texttt{ps}. RAM utilisation of the offloading process is monitored using \texttt{/proc/PID/smaps} file for the parameters $P_{5}$ and $P_{13}$. Disk utilisation of the system, denoted by $P_{3}$ and $P_{11}$ are obtained using \texttt{iotop} utility. Disk throughput of the offloading process is obtained by monitoring \texttt{/proc/PID/io} to record the number of bytes written to and read from disk for parameters $P_{6}$ and $P_{12}$. Network utilisation of the offloading process is obtained using the \textit{nethogs} tool for parameters $P_{7}$ - $P_{8}$ and $P_{15}$ - $P_{16}$
The values for offline parameters (i) $P_{17}$ is the size of the offloaded container, (ii) $P_{18}$ - $P_{23}$, are obtained from the settings defined for the Cloud and Fog VMs, and (iii) $P_{24}$ and $P_{25}$ are acquired during network configurations using \texttt{tc}.

To simulate the varying availability of CPU, memory and hard disk resources in the experimental environment, the CPU, memory and I/O stress were gradually increased for different experimental runs using \texttt{stress-ng}\footnote{https://manpages.ubuntu.com/manpages/artful/man1/stress-ng.1.html}; CPU stress was increased by 10\%, memory stress on the Cloud VM by units of 1GB until 75\% of capacity and for the Fog VM by units of 512MB until 75\% of capacity, and disk stress on the Cloud VM by units of 4GB until 75\% of capacity and for the Fog VM by units of 2GB until 75\% of capacity.

The data values using the estimation methods (of Section~\ref{sec:estimationmodels}) are used to build the model for estimating $t_{offload}$.

In the \textit{Export and Import} technique, the container state is saved using the Docker export process as a \texttt{.tar} file on the Cloud VM. The tar file is transferred to the Fog server using the scp protocol. When the tar file is received on the Fog VM, \texttt{inotify} loads the image into the local registry from which a container is booted up. During the offloading process, the values of the runtime and offline parameters are collected similar to the Save and Load technique presented above. 

In the \textit{Push and Pull} technique, the container state is pushed to a central repository on the Cloud VM using the Docker push process. On the Fog VM, the container state is pulled using the Docker pull process and a container is started from the loaded image. During offloading, the values of the runtime and offline parameters are collected same as above. 

\textit{Stateful technique}: For \textit{CRIU-based live migration}, the container state is stopped and dumped on the shared directory of the Cloud VM. \texttt{inotify} triggers the process of restoring the container state on the Fog VM. During the offloading process, the values of the runtime and offline parameters are collected similar to the discussion above.  

The dataset that is generated from all the parameters for the prediction models of the four offloading techniques consists of 5,700 instances across the two experimental platforms for the Cloud-to-Fog and Fog-to-Cloud based offload scenarios for varying combinations of the offline parameters. Since the runtime parameters are collected at one-second intervals each Cloud-to-Fog and Fog-to-Cloud offload scenario for a given set of offload parameters generates a large number of intermediate instances (on an average 60 intermediate instances). The runtime values are averaged to obtain the aggregate instances. A total of ($25 \times 5,700 \times 60 = 8,550,000$) data points are used. The experiments were repeated five times and a total of 42,750,000 data points were collected.

\subsection{Results}
The experimental results for the Cloud-to-Fog and Fog-to-Cloud offloading scenarios are presented in this section. The goal is to demonstrate the feasibility of estimating the time to offload in both scenarios. The results are presented for three metrics, namely (i) $R^2$, also known as the coefficient of determination, highlights how much of the observed variations by the regression model can be explained by the model's input. In other words, a higher value indicates a better fit of the regression model to the inputs, (ii) Mean Absolute Percentage Error (MAPE) is the average of percentage errors. A lower value indicates that the model estimates the offload time with a higher accuracy, and (iii) Mean Absolute Error (MAE) is the average error of the model. A lower value indicates that the estimation error of the model is low.  

Two validation approaches are employed in each of the offloading scenarios. The first is a train-test validation approach in which 70\% of the data is used for training and the remaining 30\% of the data is used for testing.
This approach is data-driven and requires that operational data from the infrastructure is gathered at a relatively high frequency. The second validation approach is k-fold cross validation, where $k = 10$. The results are presented for each validation approach in both offload scenarios considered in this paper. 

\subsubsection{Cloud-to-Fog offload}
\begin{figure}[t]
\begin{center}
	\subfloat[$R^2$ of the prediction models]
	{\label{fig:r2-train-cloud}
	\includegraphics[width=0.495\textwidth]
	{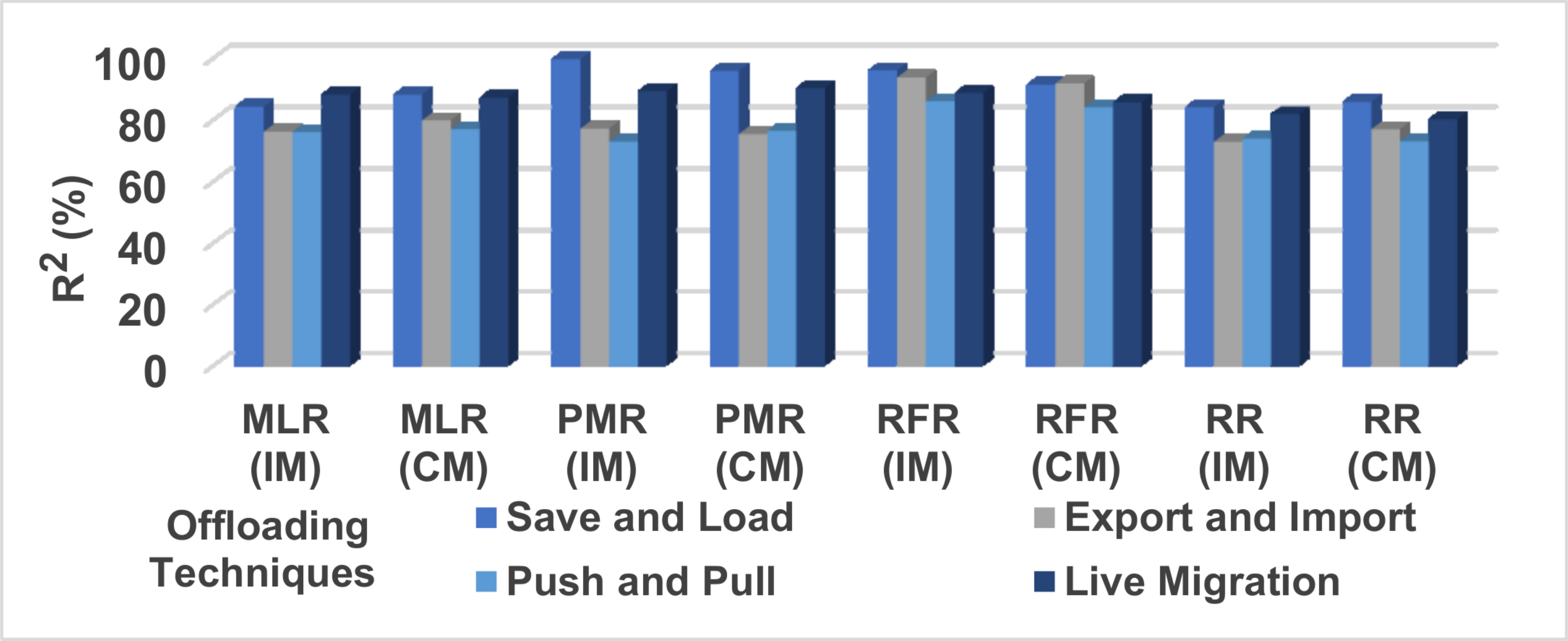}}
	
	\subfloat[Mean Absolute Error (MAE) of the prediction models]
	{\label{fig:mae-train-cloud}
	\includegraphics[width=0.495\textwidth]
	{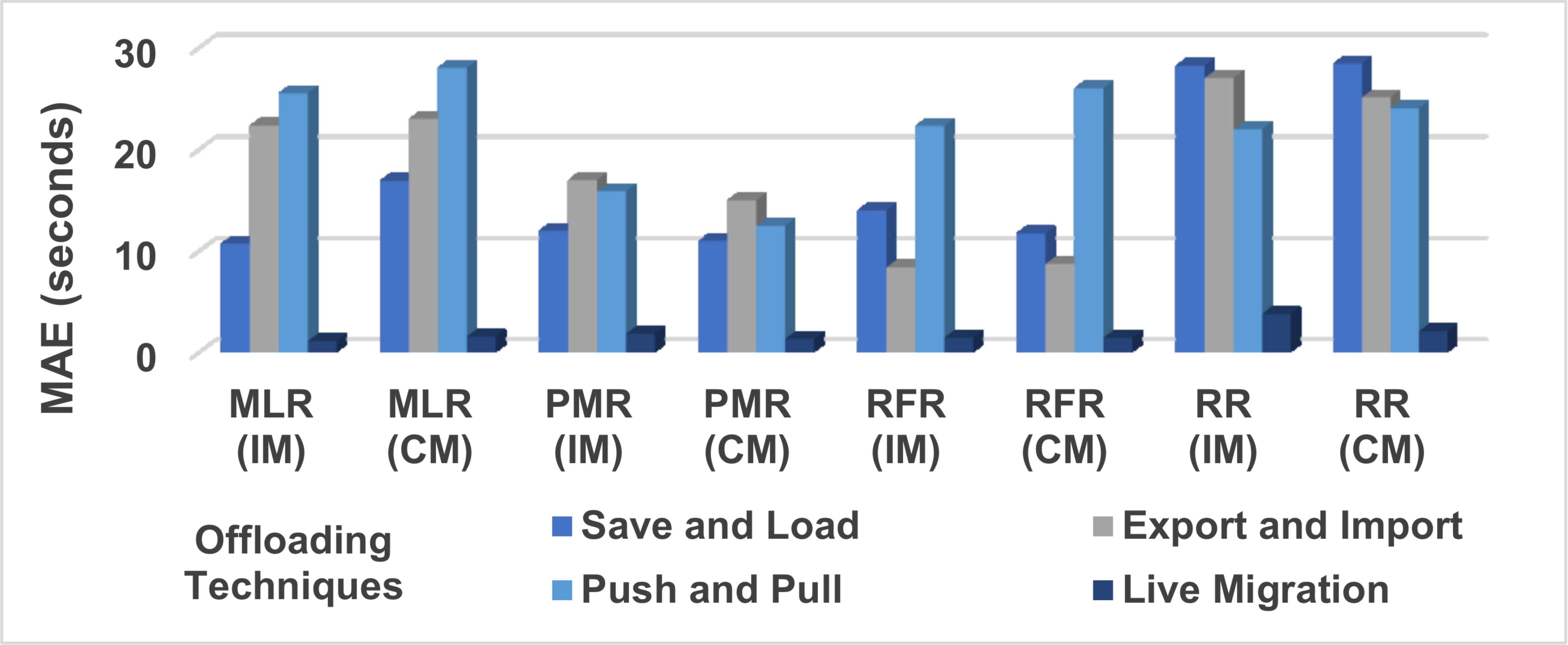}}
	
	\subfloat[Mean Absolute Percentage Error (MAPE) of the prediction models]
	{\label{fig:mpe-train-cloud}
	\includegraphics[width=0.495\textwidth]
	{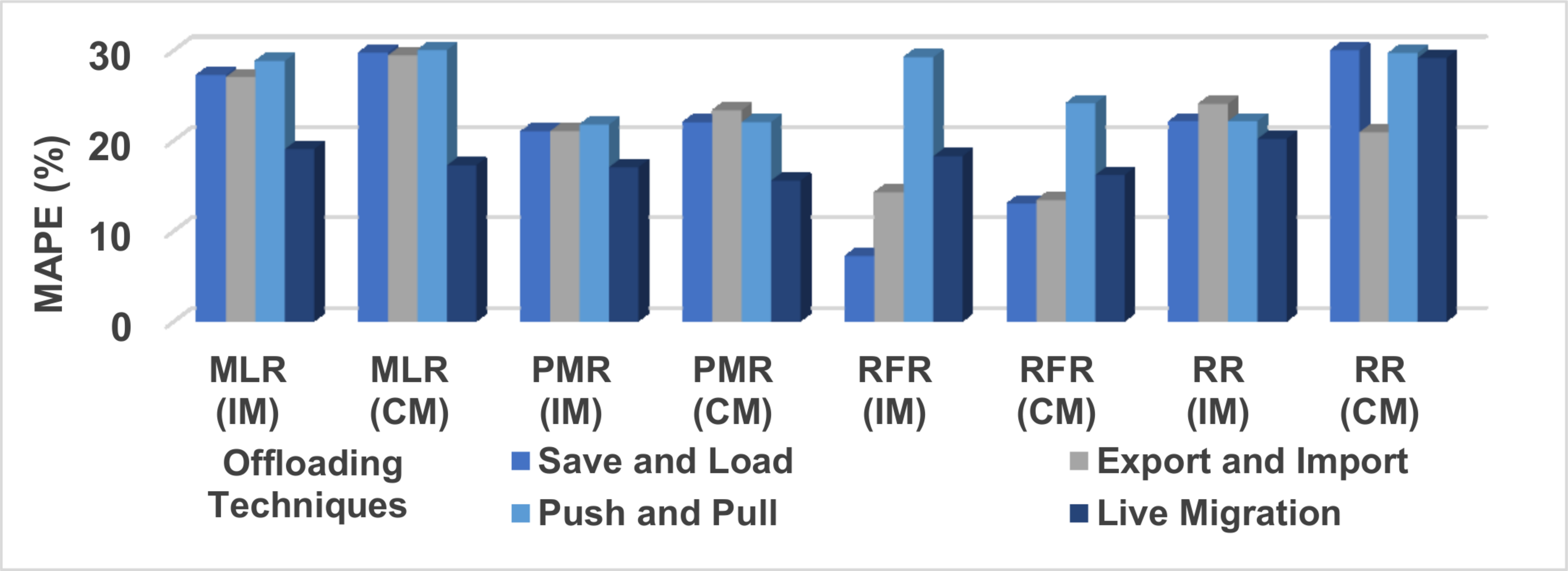}}
	
\end{center}
\caption{Train-test validation approach for Cloud-to-Fog offload}
\label{fig:train-cloud}
\end{figure}

\begin{figure}[htp]
\begin{center}
	\subfloat[$R^2$ of the prediction models]
	{\label{fig:r2-10kfold-cloud}
	\includegraphics[width=0.495\textwidth]
	{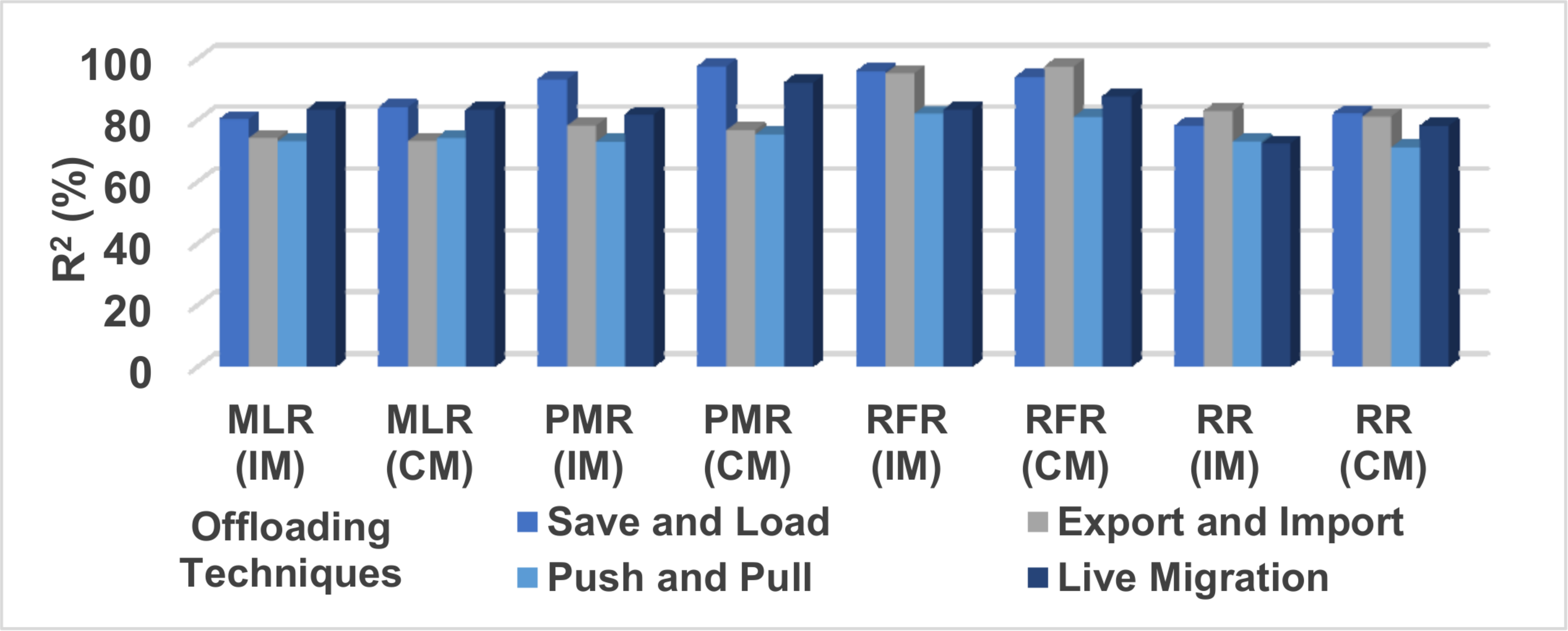}}
	 
	\subfloat[Mean Absolute Error (MAE) of the prediction models]
	{\label{fig:mae-10kfold-cloud}
	\includegraphics[width=0.495\textwidth]
	{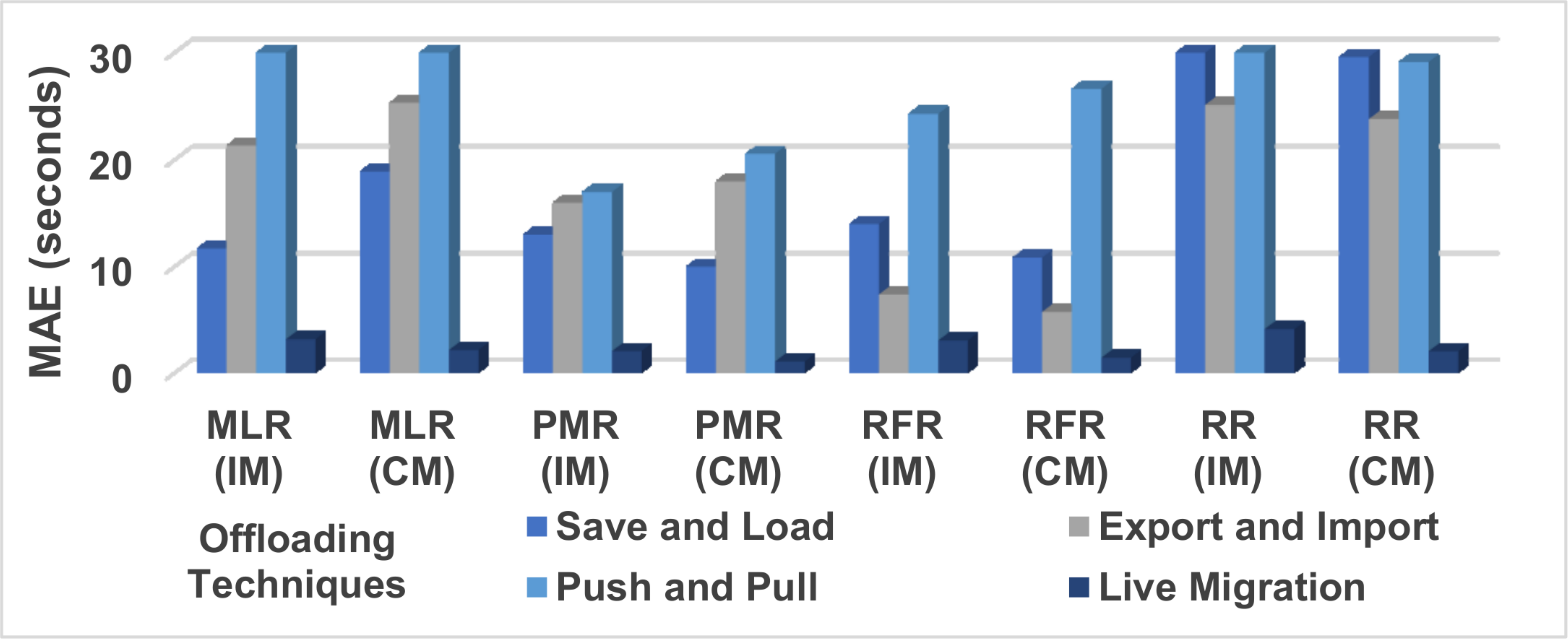}}
	
	\subfloat[Mean Absolute Percentage Error (MAPE) of the prediction models]
	{\label{fig:mpe-10kfold-cloud}
	\includegraphics[width=0.495\textwidth]
	{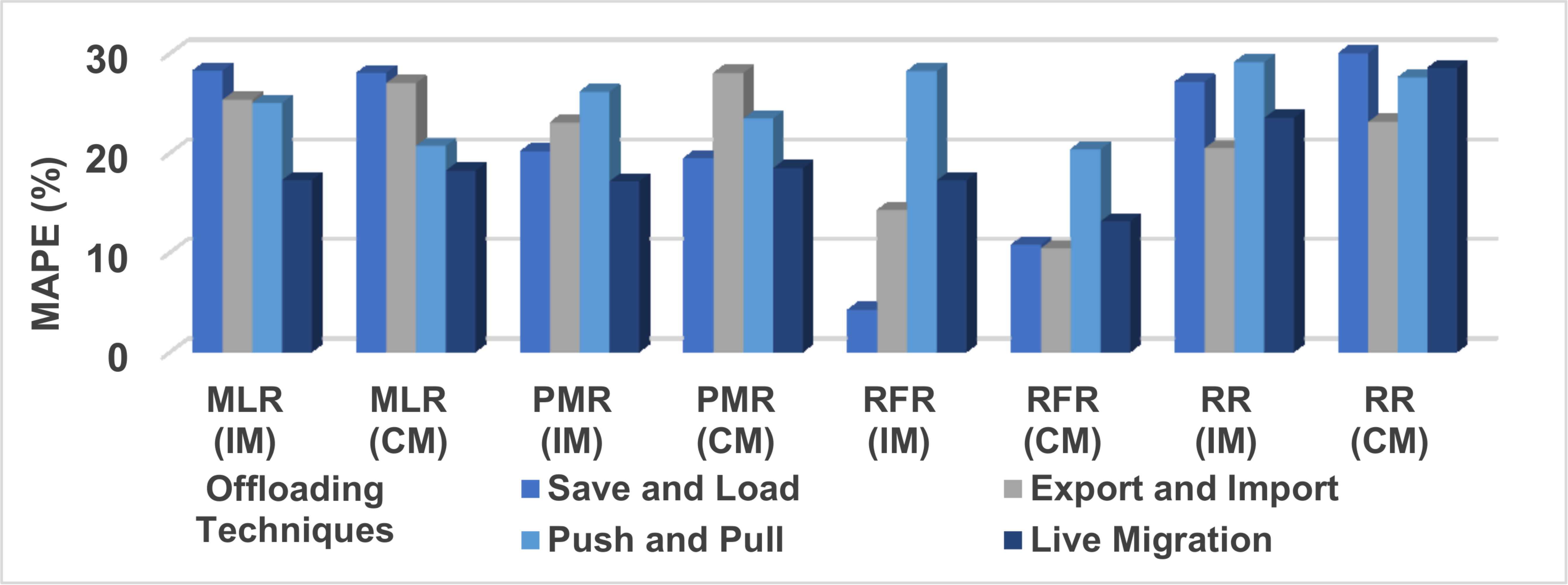}}
	
\end{center}
\caption{k-fold validation approach for Cloud-to-Fog offload}
\label{fig:10kfold-cloud}
\end{figure}

Results for the train-test validation and k-fold validation approaches are shown in Figure~\ref{fig:train-cloud} and Figure~\ref{fig:10kfold-cloud}, respectively. 

Figure~\ref{fig:r2-train-cloud} (using the train-test validation approach) and Figure~\ref{fig:r2-10kfold-cloud} (using the k-fold validation approach) show the degree of influence ($R^2$) of the chosen runtime and offline parameters (as shown in Table~\ref{tab:tab1}) on $t_{offload}$ using the four machine learning approaches when employing individual and collective models (IM and CM). Across all machine learning approaches and for both IM and CM, RFR provides the highest $R^2$. It is observed that the Push and Pull technique has the lowest $R^2$. In this technique, the central repository is accessed for uploading and downloading the container image over the Internet, and the model does not capture the associated uncertainty.
Based on $R^2$ of the estimation models, it can be concluded that the chosen parameters influence $t_{offload}$.

Figure~\ref{fig:mae-train-cloud} and Figure~\ref{fig:mae-10kfold-cloud} show the Mean Absolute Error (MAE) in the estimations using multiple validation approaches; lower value indicates a more efficient prediction in terms of the standard deviation. The RFR has the lowest MAE with the exception of the Push and Pull technique across both CM and IM. This is expected since RFR is known for more efficient feature selection and fitting. 

Figure~\ref{fig:mpe-train-cloud} and Figure~\ref{fig:mpe-10kfold-cloud} show the Mean Absolute Percentage Error (MAPE) in the estimations; lower value indicates a more efficient model and RFR has the lowest MAPE. 

Across all metrics of evaluation, RFR is a superior estimation model. The results for MAE and MAPE are surprising given that although RR applies a penalty for reducing the errors, it is less accurate in its prediction.  

\subsubsection{Fog-to-Cloud offload}

\begin{figure}[htp]
\begin{center}
	\subfloat[$R^2$ of the prediction models]
	{\label{fig:r2-train-fog}
	\includegraphics[width=0.495\textwidth]
	{sections/images/r2-train-fog.pdf}}
	
	\subfloat[Mean Absolute Error (MAE) of the prediction models]
	{\label{fig:mae-train-fog}
	\includegraphics[width=0.495\textwidth]
	{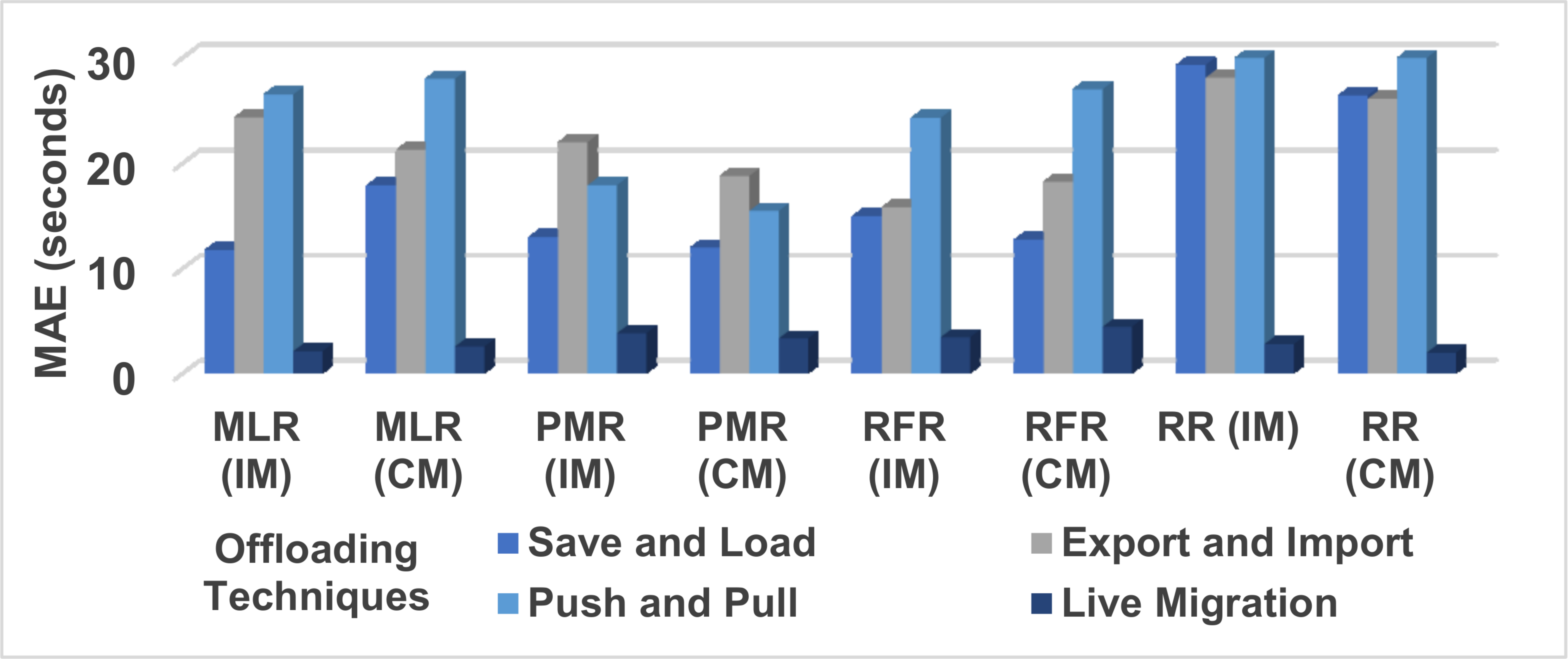}}
	
	\subfloat[Mean Absolute Percentage Error (MAPE) of the prediction models]
	{\label{fig:mpe-train-fog}
	\includegraphics[width=0.495\textwidth]
	{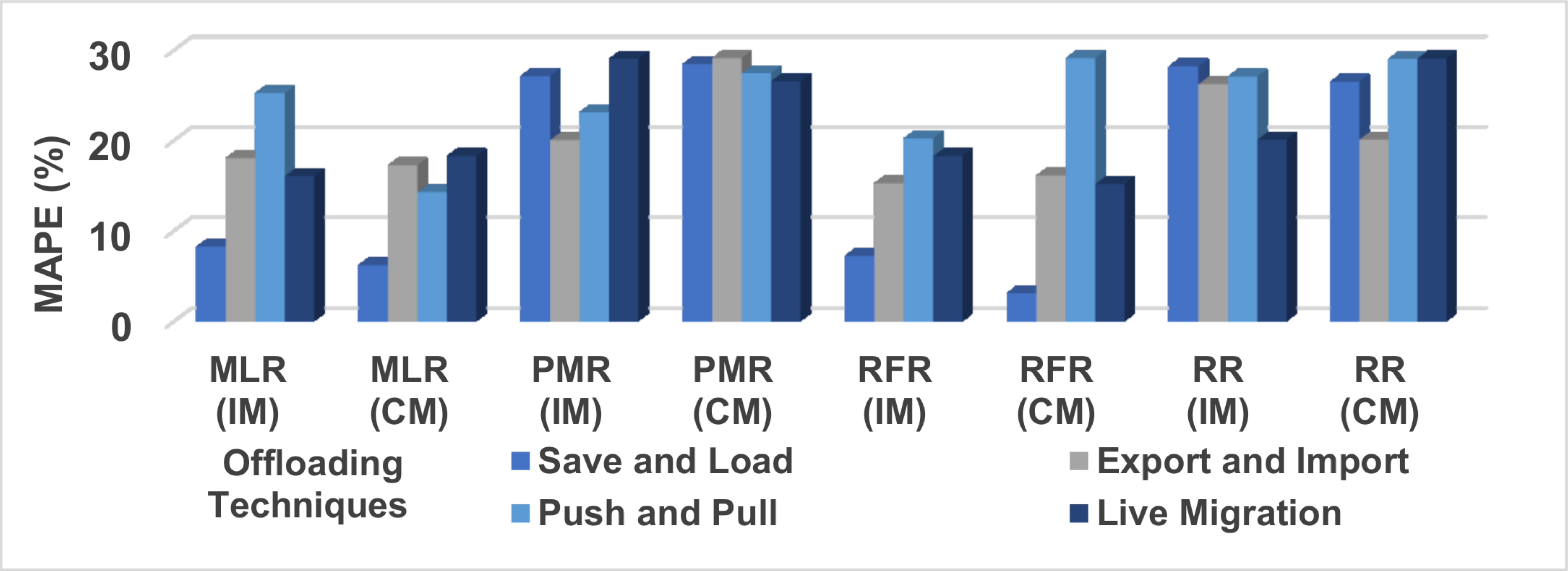}}
	
\end{center}
\caption{Train-test validation approach for Fog-to-Cloud offload}
\label{fig:train-fog}
\end{figure}

\begin{figure}[htp]
\begin{center}
	\subfloat[$R^2$ of the prediction models]
	{\label{fig:r2-10kfold-fog}
	\includegraphics[width=0.495\textwidth]
	{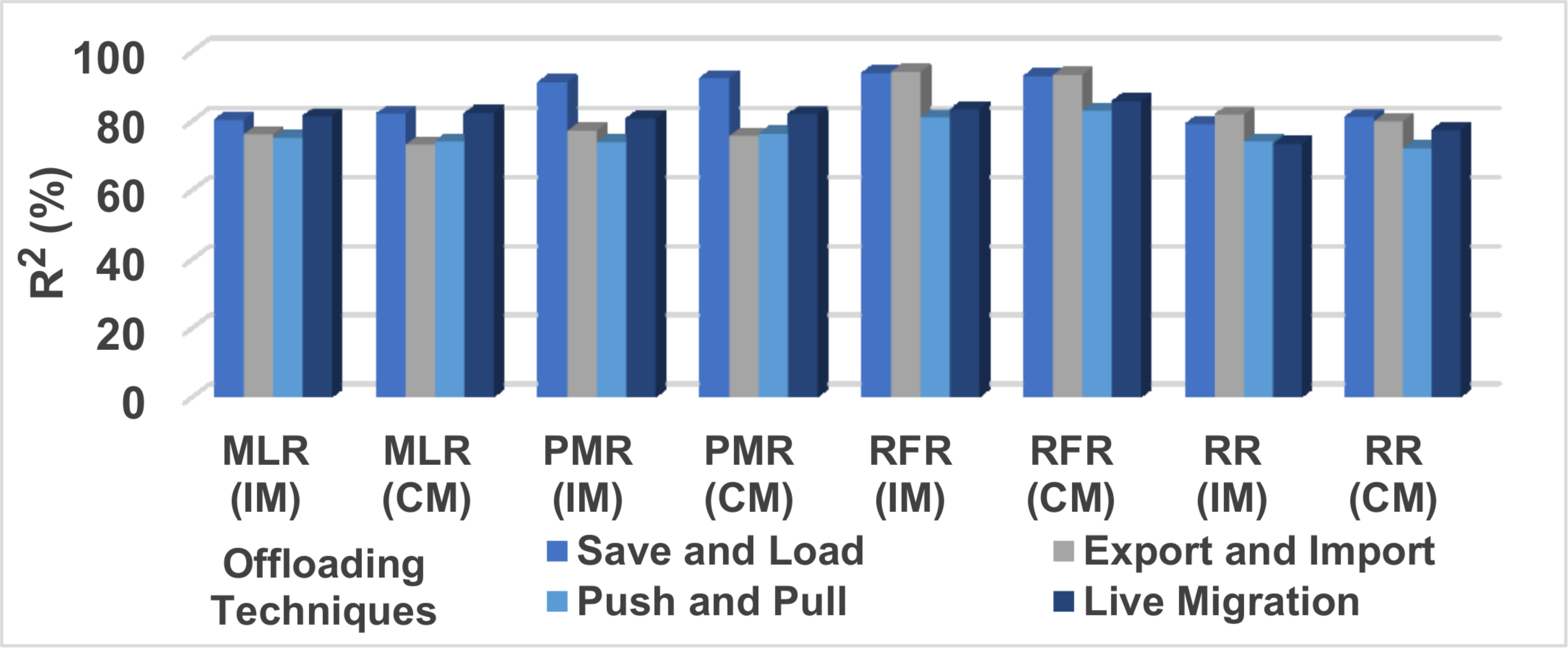}}
	
	\subfloat[Mean Absolute Error (MAE) of the prediction models]
	{\label{fig:mae-10kfold-fog}
	\includegraphics[width=0.495\textwidth]
	{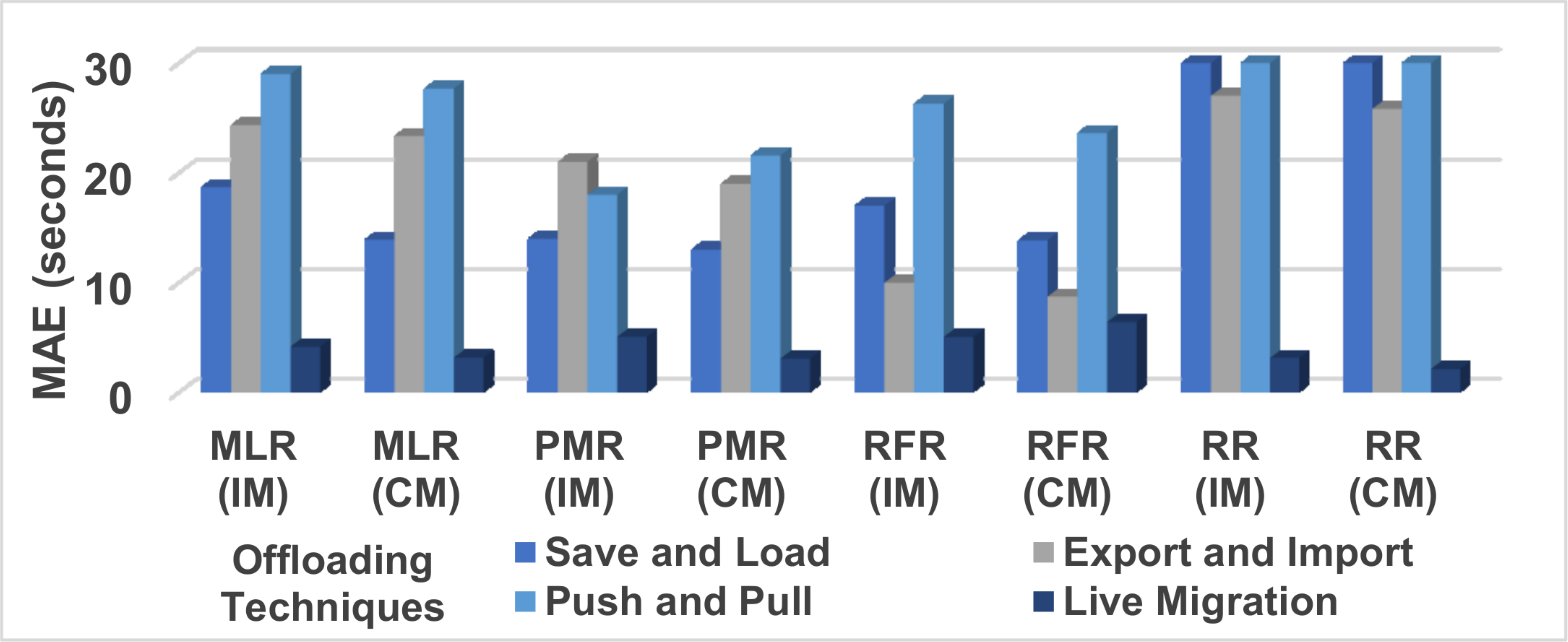}}
	
	\subfloat[Mean Absolute Percentage Error (MAPE) of the prediction models]
	{\label{fig:mpe-10kfold-fog}
	\includegraphics[width=0.495\textwidth]
	{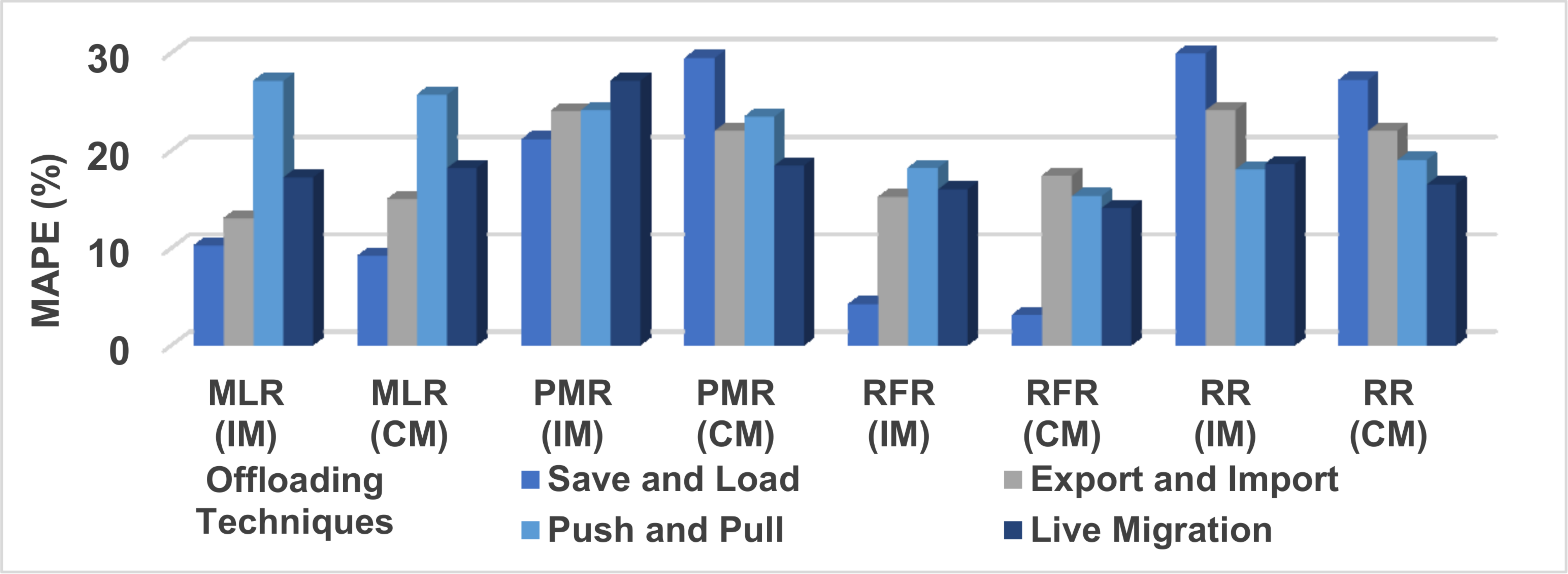}}
	
\end{center}
\caption{k-fold validation approach for Fog-to-Cloud offload}
\label{fig:10kfold-fog}
\end{figure}

The aim of presenting results for offloading from the Fog-to-Cloud is to examine whether any additional factors influence the cost of offloading, such as the cost of data preparation on the Fog due to its limited resources when compared to the cloud.
The results for $R^2$ obtained for the Fog-to-Cloud offload scenario are similar to the Cloud-to-Fog scenario (Figure~\ref{fig:r2-train-fog} and Figure~\ref{fig:r2-10kfold-fog} for the two validation approaches). The observations are similar to those noted for the Cloud-to-Fog scenario. RFRR emerges as a superior model in relation to MAE (except for Push and Pull; Figure~\ref{fig:mae-train-fog} and Figure~\ref{fig:mae-10kfold-fog}) and MAPE (Figure~\ref{fig:mpe-train-fog} and Figure~\ref{fig:mpe-10kfold-fog}).It is noted that the offloading results obtained from Cloud-to-Fog offloading and vice versa in our experimental setup are similar with small variations that do not affect the offloading model.

\subsection{Summary}
The following summarises the experimental results:
    
    (i) RFR has superior performance compared to the MLR, PMR and RR. These approaches with the runtime and offline parameters were considered to address Q1. 

    (ii) With respect to $R^2$, it is noted that: (a) for the Cloud-to-Fog scenario between 72\%--99\% of the chosen input parameters (runtime and offline) influence $t_{offload}$, and (b) for the Fog-to-Cloud scenario between 71\%--95\% of the parameters affect the offload time. This addresses Q2 posed initially by determining the influence of input parameters on offloading.

    (iii) MAE as low as 1.4 seconds is obtained for Cloud-to-Fog and 1.1 seconds for Fog-to-Cloud. The lowest MAPE for Cloud-to-Fog is 4.21\% and for Fog-to-Cloud is 3.19\%. The MAE and MAPE measures provide insight into the estimation accuracy of different models to address Q3.

    (iv) It is known that the CM approach is easier to use in practice than the IM approach (since a collection of models are required). However, it is demonstrated that accuracy obtained using the CM method is comparable with the IM method.

\section{Related Work}
\label{sec:relatedwork}
Fog offloading is necessary to meet the overall Quality-of-Service requirements of an application. This section considers three relevant areas of Fog offloading, namely the scenarios, approaches, and the decisions to be made for offloading. 


\subsubsection{Offloading Scenarios}
Offloading offers the benefits of meeting the computational requirements of an application and latency demands, load balancing, and managing energy consumed~\cite{management2018,challenge2016}. Section~\ref{sec:techniques} highlighted three offloading scenarios, namely Cloud-to-Fog, Fog-to-Cloud, and Device-to-Fog with the offloading approaches.

\textit{Cloud-to-Fog}: Current techniques to partition a monolithic application (i.e. not designed as a micro-service based application) are manual~\cite{wang2017enorm}. Multiple approaches are adopted to make decisions on Fog placement. Examples include a heuristic-based scheduling algorithm to balance makespan and the monetary cost of Cloud resources has been proposed~\cite{pham2016towards}. 

\textit{Fog-to-Cloud}: Fog resources are anticipated to be hardware constrained and geographically dispersed when compared to a Cloud data center~\cite{management2018}. Therefore, services may not be able to easily scale across the Fog if they have significant compute or storage requirements. Thus, a Fog service may need to be offloaded back to the Cloud. One strategy is to employ an analytical queuing model that is based on the LRU filter~\cite{popularitybased}.

\textit{Device-to-Fog}: Osmotic computing provides an architecture to deal with Device-to-Fog offloading~\cite{osmoticcomputing-01}. Major concerns that need to be addressed while offloading from user devices or sensors onto the Fog include minimising delays~\cite{yousefpour2018reducing}, data transferred to the Cloud~\cite{yu2017novel}, and communication overheads~\cite{hou2019fog}. 

\subsubsection{Offloading Approaches}
Virtual Machine (VM) and containers are the two main approaches considered in the literature for offloading services.  
VMs when compared to containers have larger overheads and are generally used in resource abundant environments, such as the Cloud. In the context of limited compute and storage resources as seen in the Fog, containers may be more appropriate for offloading services. Nonetheless, VM placement in the Fog by taking user mobility into account using integer linear programming has been proposed~\cite{gonccalves2018proactive}.

The majority of research in Fog computing that focuses on offloading takes containers into account. 
Service hand-off for offloading services between Fog resources is proposed~\cite{ma2017efficient} and live container migration using CRIU is considered~\cite{nadgowda2017voyager}.

\subsubsection{Offloading Decision}
Three aspects relevant to the decision-making process for Fog offloading need to be considered: (i) \textit{Which services need to be offloaded} - an application may be a composition of multiple services and not all services would equally benefit from being offloaded. Therefore, the subset of services that can benefit from offloading needs to be identified. (ii) \textit{Where to offload services} - multiple edge resources may be available within the same geographic region (for example, a street or city) to offload services. The most appropriate resource that can execute the service to meet all service objectives needs to be determined. (iii) \textit{How to offload services} - there are multiple techniques for offloading services. For example, when using containers the most appropriate approach (for example with least overhead) needs to be chosen from multiple stateless and stateful techniques (Section~\ref{sec:techniques}). 

\textit{Which services to offload}: 
Service selection has been formulated using Mixed Integer Non Linear Programming and solved using linearisation and genetic algorithms~\cite{OptimizedPlacement}. Search-based algorithms have also been considered to determine eligible services of an IoT application by taking into account the hardware capacity of Fog resources~\cite{mahmud2018latency}. 
A QoS-aware model is proposed for deploying IoT services to Fog infrastructure. The model considers the available infrastructure (latency and bandwidth), interactions among software components, and business policies during the selection of services~\cite{brogi2017qos}.

\textit{Where to Offload}:
Fog nodes are selected using multiple criteria mixed-integer linear programming~\cite{da2019location} and Markov-Decision-Process~\cite{follow-me2019} approaches. The FOLO framework performs dynamic task allocation in vehicular Fog computing that takes the resource constraints of Fog resources~\cite{zhu2018fog}. 

\textit{How to Offload}:
Low end-to-end latency for mobile users is achieved by a service hand-off mechanism using live container migration that is achieved using CRIU~\cite{ma2017efficient}. Cloud4IoT is a platform that performs vertical and horizontal migration of IoT services using a stateless approach, namely Push and Pull~\cite{dupont2017edge}. A Cloud to Fog based model to predict offloading time for stateless container technology has been proposed using machine learning techniques~\cite{majeed2019performance}. Live container migration techniques, namely Pre Copy, Post Copy, and Hybrid have been considered in the context of Fog computing environments~\cite{puliafito2019container}.
Stateful and stateless migration with low down time has been investigated~\cite{liveMigration18,ma2017efficient,puliafito2019container}.
 
The research above presents techniques for optimising offloading, service placement in the Fog and selecting the offloading approach. The research in this paper aims to characterise offloading by estimating the time taken to offload when multiple stateless and stateful techniques are employed.

\section{Conclusions}
\label{sec:conclusions}
This paper proposes the design and implementation of methods for estimating the offload time using containers in Fog computing. Four container-based offloading technique (three stateless techniques, namely Save and Load, Export and Import and Push and Pull, and one stateful technique, namely CRIU-based Live Migration) are investigated. Four estimation models based on Multivariate Linear Regression, Polynomial Multivariate Regression, Random Forest Regression and Ridge Regression are considered to estimate the offload time. A catalogue of 25 parameters collected at the system and process levels during runtime and offline are used as input to the models. Two estimation methods, namely using a collective model and individual models are proposed. Experimental studies are pursued on two Cloud-Fog lab-based platforms from which 42 million data points are obtained. The results are presented in relation to: (i) the influence of the input parameters on the estimation models measured by the coefficient of determination for regression, and (ii) the accuracy of estimation measured by the mean absolute error and mean absolute percentage error. It is noted that the Random Forest Regression (RFR) has superior performance compared to the other approaches. No specific trend was observed in relation to the collective or individual models. The input parameters are appropriate for the estimation models since up to a 99\% influence is observed on the offload time. Moreover, it is noted that the RFR can estimate the offload time with less than a 10\% mean absolute percentage error. 


In the future, an integrated decision-making approach that considers: (i) which services of Directed Acyclic Graph based applications should be offloaded, and (ii) given multiple Fog resources where should they be offloaded will be developed.

\section*{Acknowledgment}
The first author is funded by a Schlumberger Scholarship.

\bibliographystyle{IEEEtran}  
\bibliography{references}

\end{document}